\documentclass[10pt,twocolumn,twoside]{IEEEtran}
\ifCLASSINFOpdf
  \usepackage[pdftex]{graphicx}
  \graphicspath{{./figures/}}
  \DeclareGraphicsExtensions{.pdf,.jpeg,.png}
\else
  \usepackage[dvips]{graphicx}
  \graphicspath{{./figures/}}
  \DeclareGraphicsExtensions{.eps}
\fi
\ifCLASSINFOpdf
\usepackage[pdftex,pdfborder={0 0 0}]{hyperref} 
\else
\usepackage[dvips,pdfborder={0 0 0}]{hyperref} 
\fi

\usepackage[cmex10,tbtags]{amsmath}
\usepackage{amssymb}

\allowdisplaybreaks

\usepackage{algpseudocode}

  \usepackage[caption=false,font=footnotesize]{subfig}
\usepackage{float}

\usepackage{fixltx2e}

\usepackage{cite}

\newcommand{\Amat}{{\boldsymbol{A}}}
\newcommand{\bvec}{{\boldsymbol{b}}}
\newcommand{\bvecu}{\bvec}
\newcommand{\dbar}{{\bar{d}}}
\newcommand{\Gmat}{{\boldsymbol{G}}}
\newcommand{\Imat}{{\boldsymbol{I}}}
\newcommand{\calL}{{\mathcal{L}}}
\newcommand{\sbar}{{\bar{s}}}
\newcommand{\svec}{{\boldsymbol{s}}}
\newcommand{\ubar}{{\bar{u}}}
\newcommand{\uvec}{{\boldsymbol{u}}}

\newcommand{\xvec}{{\boldsymbol{x}}}
\newcommand{\yvec}{{\boldsymbol{y}}}
\newcommand{\zvec}{{\boldsymbol{z}}}

\newcommand{\muu}{\mu}

\newcommand{\ADMM}{{\text{ADMM}}}
\newcommand{\AL}[2]{\calL_A(#1;#2)}
\DeclareMathOperator*{\argmin}{arg\,min}
\newcommand{\Complex}{{\mathbb{C}}}
\newcommand{\defeq}{{\overset{\scriptscriptstyle\Delta}{=}}}
\newcommand{\FISTA}{{\text{FISTA}}}
\newcommand{\fnfun}[1]{{f_-(#1)}}
\newcommand{\fpfun}[1]{{f_+(#1)}}
\newcommand{\hfun}[2]{{h(#1;#2)}}
\newcommand{\hnfun}[2]{{h_-(#1;#2)}}
\newcommand{\hpfun}[2]{{h_+(#1;#2)}}
\renewcommand{\Im}[1]{{\mathcal{I}m\{#1\}}}
\newcommand{\mm}{{\text{mm}}}
\newcommand{\logl}[2]{{\ell(#1;#2)}}
\newcommand{\nphase}[1]{{e^{-\imath\angle #1}}}
\newcommand{\phase}[1]{{e^{\imath\angle #1}}}
\newcommand{\phifun}[3]{{\phi(#1;#2,#3)}}
\newcommand{\phinfun}[3]{{\phi_-(#1;#2,#3)}}

\renewcommand{\Re}[1]{{\mathcal{R}e\{#1\}}}
\newcommand{\Real}{{\mathbb{R}}}
\DeclareMathOperator{\rootop}{root}
\newcommand{\soft}[2]{{\mathrm{soft}(#1;#2)}}
\newcommand{\subjto}{\ {\text{s.t.}}\ }
\newcommand{\zerovec}{{\boldsymbol{0}}}

\floatstyle{ruled}
\newfloat{algorithmDSW}{tbp}{loa}
\floatname{algorithmDSW}{Algorithm}
\algnewcommand{\Break}{{\bf break}}

\begin{document}
%
\title{Undersampled Phase Retrieval with Outliers}
%
%
%

\author{Daniel~S.~Weller,~\IEEEmembership{Member,~IEEE,}
        Ayelet~Pnueli,
        Gilad~Divon,
        Ori~Radzyner,\\
        Yonina~C.~Eldar,~\IEEEmembership{Fellow,~IEEE,}
        and~Jeffrey~A.~Fessler,~\IEEEmembership{Fellow,~IEEE}
\thanks{DSW was funded by National Institutes of Health (NIH) grant F32 EB015914.
JAF is funded in part by NIH grant P01 CA87634
and an equipment donation from Intel. 
YCE is funded in part by Israel Science Foundation Grant 170/10, 
SRC, 
and Intel Collaborative Research Institute for Computational Intelligence.}%
\thanks{DSW is with the Charles L.\ Brown Department of Electrical and Computer Engineering, University of Virginia, Charlottesville, VA 22904 USA (email:
dweller@virginia.edu). AP was with, and GD, OR, and YCE are with the Electrical Engineering Department, Technion, Israel Institute of Technology, Haifa 32000, Israel (emails: ayelet.pnueli@gmail.com, giladd44@gmail.com, radzy@campus.technion.ac.il, yonina@ee.technion.ac.il). JAF is with the Department of Electrical Engineering and Computer Science, University of Michigan, Ann Arbor, MI 48109 USA (email: fessler@umich.edu).}
}

%
%

\markboth{Submitted to IEEE Transactions on Computational Imaging}%
{Weller \MakeLowercase{\textit{et al.}}: Undersampled Phase Retrieval with Outliers}
%



\maketitle

\begin{abstract}
We propose a general framework 
for reconstructing transform-sparse images 
from undersampled (squared)-magnitude data 
corrupted with outliers.
This framework is implemented using 
a multi-layered approach, 
combining multiple initializations 
(to address the nonconvexity of the phase retrieval problem), 
repeated minimization of a convex majorizer (surrogate for a nonconvex objective function), 
and iterative optimization 
using the alternating directions method of multipliers.
Exploiting the generality
of this framework, 
we investigate using a Laplace measurement noise model 
better adapted to outliers 
present in the data
than the conventional Gaussian noise model. 
Using simulations,
we explore the sensitivity 
of the method to both the regularization 
and penalty parameters. 
We include 1D Monte Carlo 
and 2D image reconstruction comparisons 
with alternative phase retrieval algorithms. 
The results suggest the proposed method, 
with the Laplace noise model, 
both increases the likelihood 
of correct support recovery 
and reduces the mean squared error 
from measurements containing outliers.
We also describe exciting extensions 
made possible by the generality of the proposed framework, 
including regularization using analysis-form sparsity priors 
that are incompatible 
with many existing approaches.
\end{abstract}


\begin{center} \bfseries EDICS Categories: CIF-SBR, CIF-SBI, CIF-OBI \end{center}
%
\IEEEpeerreviewmaketitle

\section{Introduction}


\IEEEPARstart{P}{hase} retrieval~\cite{fienup:13:pra,shechtman:14:prw,klibanov:95:tpr} 
refers to the problem 
of recovering a signal 
or image 
from magnitude-only 
measurements 
of a transform 
of that signal. 
This problem 
appears 
in crystallography~\cite{sayre:52:sio,millane:90:pri,hauptman:91:tpp,harrison:93:ppi},
optical imaging~\cite{walther:63:tqo}, 
astronomical imaging~\cite{fienup:87:pra}, 
and other areas~\cite{miao:99:etm,balan:06:osr,setsompop:08:mls,chai:11:aiu,latychevskaia:11:nfd}.

Phase retrieval 
is inherently ill-posed, 
as many signals 
may share the same 
magnitude spectrum~\cite{oppenheim:81:tio}. 
To address this issue, 
existing phase retrieval algorithms 
incorporate different sources 
of prior information. 
The Gerchberg-Saxton error reduction method~\cite{gerchberg:72:apa} 
of alternating projections
uses magnitude information 
about both an image 
and its Fourier spectrum.
Fienup's 
hybrid input-output (HIO) algorithm~\cite{fienup:78:roa,fienup:82:pra}
generalizes the image domain 
projection 
of error reduction 
to other constraints such as 
image boundary 
and support information~\cite{hayes:83:rpr,fienup:86:pru,fienup:87:roa}.
More recently, 
the alternating projections framework~\cite{bauschke:03:hpr} 
has been extended to sparse reconstruction~\cite{ohlsson:14:ocf,ranieri:13:prf,eldar:14:prs}; 
examples include compressive phase retrieval~\cite{moravec:07:cpr,schniter:12:cpr}
and the sparse Fienup method~\cite{mukherjee:12:aia}. 
Other formulations forgo the HIO framework. 
One method uses rough phase estimates~\cite{osherovich:11:afp} 
to dramatically improve reconstruction quality.
Another uses a matrix lifting scheme~\cite{candes:13:pea,candes:13:prv}
to construct a semidefinite relaxation 
of the phase retrieval problem~\cite{demanet:13:crf}, 
which may be combined 
with sparsity-promoting regularization~\cite{shechtman:11:sbs,ohlsson:12:cpr,li:13:ssr,candes:13:pea,waldspurger:13:prm}. 
Other approaches 
employing sparsity 
for phase retrieval 
include the graph-based 
and convex optimization methods 
in~\cite{jaganathan:12:ros} 
and greedy algorithms 
like GESPAR~\cite{shechtman:14:gep}.

In addition 
to lacking phase information, 
measurements are often noisy, 
especially 
at the microscopic scales 
used in crystallography 
and optical imaging. 
Most existing methods 
either ignore measurement noise 
or impose quadratic data fit penalties. 
Our method, introduced first in~\cite{weller:14:pro},
employs a $1$-norm data fit term, 
corresponding to a Laplace noise model,
designed to improve robustness 
to outliers.
Our optimization framework 
combines a majorize-minimize algorithm 
with a nested variable-split 
and the alternating directions method of multipliers (ADMM) 
to solve the phase retrieval problem 
with a robust data fit model 
and $1$-norm sparsity-promoting regularizer.
Although the original problem is nonconvex, 
our proposed majorizer is convex and 
as tight as possible. 
While direct minimization of this majorizer 
would be combinatorially complex,
introducing an auxiliary variable enables 
efficient minimization via ADMM.
We compare our approach 
against using a conventional quadratic data fit term
within our framework, 
separating the contributions 
of the implementation 
from the proposed noise model.
We established earlier~\cite{weller:14:pro} 
that properly tuning the parameter 
for the $1$-norm sparse regularization term 
is essential for 
successful reconstruction. 
Here, we thoroughly study the parameter selection problem, 
analyzing the regularization parameter 
as well as the ADMM penalty parameter 
that affects the convergence rate 
of the ADMM component of the algorithm.

Section~\ref{sec:models} presents 
a general likelihood model 
for the phase retrieval problem.
Section~\ref{sec:majorizer} 
introduces a convex majorizer 
for the optimization problem, 
and Sec.~\ref{sec:ADMM} 
describes our solution 
to this convex subproblem 
using ADMM.
After investigating the tuning 
of the regularization 
and penalty parameters 
in Sec.~\ref{sec:parameters}, 
we present 1D Monte Carlo comparisons 
in Sec.~\ref{sec:mccomp},
and a 2D image reconstruction 
in Sec.~\ref{sec:imcomp}. 
We conclude 
with a discussion 
of the merits 
of our algorithm 
and future extensions.
Code and data are available online 
from~\url{http://people.virginia.edu/~dsw8c/sw.html}.

\section{Problem Statement}\label{sec:models}

Consider the standard phase retrieval problem, 
where a length-$N$ (complex-valued) signal $\xvec$ is reconstructed 
from $M$ squared-magnitude measurements $\yvec = [y_1,\ldots,y_M]^T$ 
of the discrete Fourier transform (DFT) of $\xvec$:
\begin{equation}
y_m = |[\Amat\xvec]_m|^2 + \nu_m,\quad m = 1,\ldots,M,
\label{eq:prconventional}
\end{equation}
where $[\Amat\xvec]_m = \sum_{n=1}^N A_{mn}x_n$ 
is the $m$th DFT coefficient, 
and $[\nu_1,\ldots,\nu_M]^T$ is a vector 
of additive white Gaussian noise.
The vector $\xvec$ 
may represent either a 1D signal 
or a higher dimensional image, 
columnized.

Our framework aims 
to minimize the negative log-likelihood function 
$\sum_{m=1}^M -\logl{y_m}{|[\Amat\xvec]_m|^q}$. 
With Gaussian noise, 
\begin{equation}
-\logl{y_m}{|[\Amat\xvec]_m|^q} \propto |y_m-|[\Amat\xvec]_m|^q|^2.
\label{eq:nLgaussian}
\end{equation} 
This formulation generalizes standard phase retrieval 
in several ways. 
First, the linear transform $\Amat$ 
can be any sensing matrix, 
not just the DFT.
Second, the system may measure
the magnitude or squared-magnitude 
of $[\Amat\xvec]_m$, 
or even more broadly, 
any power of the magnitude $|[\Amat\xvec]_m|^q$, 
for $q \geq 1$.
Third, 
the measurement noise 
no longer is strictly Gaussian.
%
%
To account for outliers in the data,
we focus on using the Laplace distribution, 
with negative log-likelihood function
\begin{equation}
-\logl{y_m}{|[\Amat\xvec]_m|^q} \propto |y_m-|[\Amat\xvec]_m|^q|.
\label{eq:nLlaplace}
\end{equation} 

Our method applies more broadly 
to log-likelihood functions 
of the form $-\logl{[\Amat\xvec]_m}{y_m} = f(\hfun{[\Amat\xvec]_m}{y_m})$, 
where $f(\cdot)$ is convex and nondecreasing 
(on $\Real_+$), 
and the data prediction error function 
$\hfun{t}{y}\ \defeq\ |y - |t|^q|$, 
with $t \in \Complex$ 
and $y \in \Real$.
For this class 
of functions,
the majorizer derived 
in Sec.~\ref{sec:majorizer} 
is convex in $\xvec$.

To resolve the ill-posedness 
of the phase retrieval problem, 
we impose a sparsity-promoting prior 
on the signal, 
using the $1$-norm 
convex relaxation $\|\xvec\|_1$. 
Throughout this work, 
we focus on image sparsity, 
or equivalently, synthesis-form sparsity, 
by appending a synthesis transform 
to the sensing matrix $\Amat$.
We seek the minimizer $\hat{\xvec}\in\Complex^N$ of
the problem 
\begin{equation}
\hat{\xvec} = \argmin_{\xvec\in\Complex^N} \Psi(\xvec)\ \defeq\ \sum_{m=1}^M f(\hfun{[\Amat\xvec]_m}{y_m}) + \beta\|\xvec\|_1,
\label{eq:Psix} 
\end{equation}
where $\beta > 0$ is 
the regularization penalty parameter.
Our algorithm aims 
to find a sparse signal $\xvec$ 
that is roughly consistent 
with the magnitude data.

Our formulation~\eqref{eq:Psix} differs 
from many of the methods described 
in the literature. 
First, the existing methods 
are not designed to accommodate the Laplace noise model, 
limiting their robustness to outliers. 
The projection-based methods, 
the semidefinite relaxations, 
and GESPAR all implicitly (via projections) 
or explicitly minimize the quadratic 
negative log-likelihood representing a Gaussian noise model.
In addition, the GESPAR 
and sparse Fienup methods 
use $0$-``norm'' sparsity, 
while we use $1$-norm sparsity-promoting regularization,
also found in the convex relaxations 
recently developed.

\section{Majorization of the Measurement Objective}\label{sec:majorizer}


The inverse problem formulation 
of phase retrieval 
is particularly difficult to solve 
because having only magnitude information 
makes the data fit term 
in the objective function $\Psi(\xvec)$ 
nonconvex.
To facilitate optimization, 
we construct 
a convex majorizer 
for $\Phi(\xvec)$.
Section~\ref{sec:ADMM} 
describes an iterative method 
for minimizing this majorizer effectively.

\subsection{Majorizing $\Psi(\xvec)$}

In general, 
a majorizer $\phi(t;s)$ 
for a function $h(t)$ 
satisfies two properties: 
$\phi(s;s) = h(s)$, 
and $\phi(t;s) \geq h(t)$, 
for all $t$.
When these properties are satisfied, 
decreasing the value 
of the majorizer 
also decreases 
the value of the original function, 
since $h(t) \leq \phi(t;s) < \phi(s;s) = h(s)$.

Returning to our framework, 
assuming $f(\cdot)$ is 
convex and nondecreasing, 
and $\phi(\cdot)$ 
is a convex function, 
then $f(\phi(\cdot))$ 
is convex~\cite{boyd:04}.
So, given a convex surrogate $\phifun{t}{s}{y}$ 
for $\hfun{t}{y}$, 
$f(\phifun{t}{s}{y})$ is convex. 
Furthermore, when $\phifun{t}{s}{y}$ majorizes $\hfun{t}{y}$, 
$f(\phifun{t}{s}{y})$ majorizes $f(\hfun{t}{y})$ as well.
To find a convex majorizer $\phifun{t}{s}{y}$, 
we first write $\hfun{t}{y}$ as
\begin{equation}
\hfun{t}{y} = \max\{\hpfun{t}{y}\ \defeq\ |t|^q - y,\ \hnfun{t}{y}\ \defeq\ y - |t|^q\}.
\end{equation}
Assuming $q \geq 1$, 
$\hpfun{t}{y}$ is already convex, 
but $\hnfun{t}{y}$ is concave. 
When $y \leq 0$, 
$\hfun{t}{y} = \hpfun{t}{y}$. 
But, whenever $y > 0$, 
$\hnfun{t}{y}$ needs to be replaced 
with a convex majorizer $\phinfun{t}{s}{y}$. 
Then, $\phifun{t}{s}{y}\ \defeq\ \max\{\hpfun{t}{y},\ \phinfun{t}{s}{y}\}$ 
is convex 
and majorizes $\hfun{t}{y}$.

\begin{figure}[!tbp]
\centering
\includegraphics[width=3in]{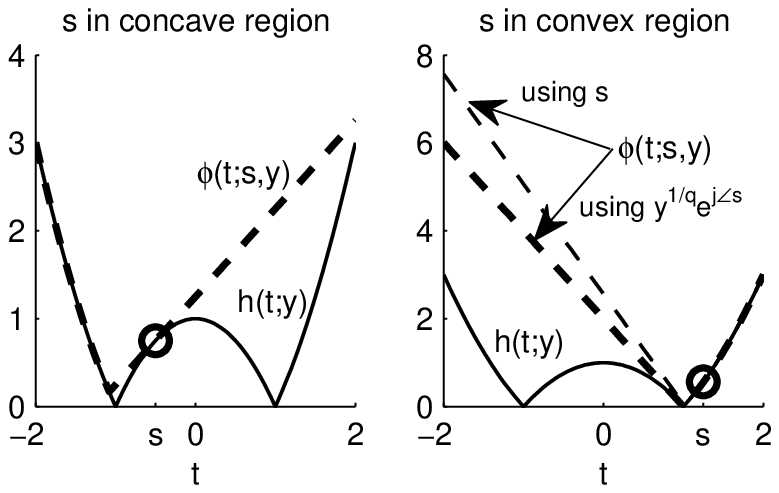}
\vspace{-0.1in}
\caption{The data fit error $\hfun{t}{y}$ (solid line) 
and the convex majorizer (surrogate) $\phifun{t}{s}{y}$ (dashed line) 
are plotted for real $t$, $y = 1$, and $q = 2$. Circles highlight 
the majorization points $s$ 
for both examples. 
In the left figure, the majorization point $s$ 
is in the concave region of $\hfun{t}{y}$, 
so the tangent plane at $s$ is used in this region. 
In the right figure, $s$ is located in the convex region 
of $\hfun{t}{y}$, 
and the tangent plane at $y^{1/q}e^{\imath\angle s}$ is used instead.}
\label{fig:surrogate_examples}
\end{figure}
Since $\hnfun{t}{y}$ is concave, 
we employ 
as a convex surrogate 
its tangent plane 
about some point $s \in \Complex$:
\begin{equation}
\begin{split}
\phinfun{t}{s}{y} &= (y - |s|^q) + (-q|s|^{q-1})\Re{\nphase{s}(t-s)}\\
&= y + (q-1)|s|^q - q|s|^{q-1}\Re{t\nphase{s}}.
\end{split}
\label{eq:phinegt}
\end{equation}
When $q = 1$, 
$\hnfun{t}{y}$ is not differentiable 
at $t = 0$, 
but our definition in~\eqref{eq:phinegt} 
is consistent with 
the tangent plane $\phinfun{t}{s}{y} = y$ 
in this context.

Since any other convex majorizer 
must lie above the tangent plane, 
\eqref{eq:phinegt}~is clearly 
tight among possible convex majorizers 
of $\hnfun{t}{y}$. 
However, 
when using $|s|^q > y$, 
we are in the convex region 
of $\hfun{t}{y}$, 
and we only need to majorize $\hfun{t}{y}$ 
in the concave region. 
In this case, 
the tangent plane 
$\sbar\ \defeq\ y^{1/q}\phase{s}$ 
still majorizes $\hnfun{t}{y}$ 
in the range of $|t|^q \leq y$. 

Our majorizer $\phifun{t}{s}{y}$ is
therefore given by
\begin{equation}
\phifun{t}{s}{y} = \begin{cases} \hpfun{t}{y}, & y \leq 0,\\
\max\{\hpfun{t}{y},\ \phinfun{t}{s}{y}\}, & 0 \leq |s|^q < y,\\
\max\{\hpfun{t}{y},\ \phinfun{t}{\sbar}{y}\}, & 0 < y \leq |s|^q.\end{cases}
\label{eq:phit}
\end{equation}
The first case occurs 
when $\hfun{t}{y}$ is already convex 
($|t|^q$ cannot be less than $y$).
The second and third cases correspond 
to $s$ 
being in the concave 
and convex regions 
of $\hfun{t}{y}$, 
respectively.
Figure~\ref{fig:surrogate_examples} portrays 
examples of the function $\hfun{t}{y}$ 
and its surrogate $\phifun{t}{s}{y}$ 
in both the second ($s$ in concave region) 
and third ($s$ in convex region) cases.
Substituting $\phifun{t}{s}{y}$ 
for $\hfun{t}{y}$ 
in the original objective 
yields our complete convex surrogate $\Phi(\xvec;\svec)$
for $\Psi(\xvec)$:
\begin{equation}
\Phi(\xvec;\svec) = \sum_{m=1}^M f(\phifun{[\Amat\xvec]_m}{s_m}{y_m}) + \beta\|\xvec\|_1.
\label{eq:Phixs}
\end{equation}

\subsection{Majorize-Minimize Algorithm}

Our proposed majorized approach to minimizing $\Psi(\xvec)$ 
in~\eqref{eq:Psix} 
repeatedly minimizes $\Phi(\xvec;\svec)$, 
using the majorize-minimize~\cite{lange:00:otu,jacobson:07:aet} scheme
shown in Algorithm~\ref{alg:mm}.
\begin{algorithmDSW}[!tbp]
\begin{algorithmic}
\Require $I_\mm$, $\epsilon_\mm$, random $\svec^0\in\Complex^M$.
\For{$i = 1:I_\mm$}
\begin{align}
\xvec^{i} &\leftarrow \argmin_\xvec \Phi(\xvec;\svec^{i-1}).\hspace{1.25in}
\label{eq:mmxupdate}\\
\svec^{i} &\leftarrow \Amat\xvec^{i}.
\label{eq:mmsupdate}
\end{align}
\If{$\|\svec^{i} - \svec^{i-1}\| < \epsilon_\mm$}
\Break
\EndIf
\EndFor
\end{algorithmic}
\caption{Majorize-minimize scheme for solving~\eqref{eq:Psix}.}
\label{alg:mm}
\end{algorithmDSW}
Although each iteration 
of this majorize-minimize method 
decreases $\Psi(\xvec)$, 
convergence 
to a minimum of $\Psi(\xvec)$ 
is not guaranteed, 
since the majorizer may get ``stuck''
at a critical point of $\Psi(\xvec)$, 
like the local maximum 
at $t = 0$.
Since the original problem 
is nonconvex, 
running the algorithm 
for multiple different initializations 
increases the chance 
of finding a global optimum
while decreasing the likelihood 
of failure 
due to stagnation.
Employing multiple initializations 
is frequently employed by other phase retrieval methods 
and when solving nonconvex problems more generally.

\section{Solving the Majorized Objective with ADMM}\label{sec:ADMM}


Jointly minimizing $M$ pairwise maximum functions 
to solve~\eqref{eq:Phixs} directly 
would be combinatorially complex.
Instead, we introduce an auxiliary vector $\uvec = \Amat\xvec$ 
to un-mix $\Amat\xvec$ 
and ensure each function 
in the summation in~\eqref{eq:Phixs} 
depends only on a single $u_m = [\uvec]_m$. 
The constrained optimization problem 
using this auxiliary variable is 
\begin{equation}
\begin{split}
\{\xvec^{i+1},\uvec\} &\leftarrow \argmin_{\xvec,\uvec} \sum_{m=1}^M f(\phifun{u_m}{s_m}{y_m}) + \beta\|\xvec\|_1,\\
&\quad\quad\ \subjto u_m = [\Amat\xvec]_m,\quad m = 1,\ldots,M.
\end{split}
\label{eq:Phixscon}
\end{equation}

We use the alternating directions method of multipliers (ADMM)~\cite{glowinski:75:slp,gabay:76:ada,eckstein:92:otd,boyd:10:doa} 
framework to solve the augmented Lagrangian form 
of this constrained problem:
\begin{equation}
\begin{split}
\AL{\xvec,\uvec}{\bvecu}\ &\defeq\ \sum_{m=1}^M f(\phifun{u_m}{s_m}{y_m}) + \beta\|\xvec\|_1\\
&\quad + \tfrac{\muu}{2}\|\Amat\xvec - \uvec + \bvecu\|_2^2,
\end{split}
\label{eq:PhixsAL}
\end{equation}
where $\bvecu\in\Complex^M$ and $\muu > 0$ are 
the scaled dual vector (Lagrange multipliers) 
and augmented Lagrangian penalty parameter, 
respectively.
Our implementation of ADMM 
in Algorithm~\ref{alg:ADMM} 
solves~\eqref{eq:PhixsAL}.
We define $d_m = [\Amat\xvec+\bvecu]_m$ 
to simplify notation 
here and in subsequent sections.
\begin{algorithmDSW}[!tbp]
\begin{algorithmic}
\Require $I_\ADMM$, $\epsilon_\ADMM$, $\xvec^0$, $\uvec^0$, $\yvec$, $\beta$, $\muu$.
\State $\bvecu^0 \leftarrow \Amat\xvec^0 - \uvec^0$.
\For{$i = 1:I_\ADMM$}
\begin{equation}
\xvec^{i} \leftarrow \argmin_\xvec \beta\|\xvec\|_1\! +\! \tfrac{\muu}{2}\|\Amat\xvec\! -\! (\uvec^{i-1}\! -\! \bvecu^{i-1})\|_2^2.\hspace{-8pt}\label{eq:admmxupdate}
\end{equation}
\For{$m = 1:M$}
\State $d_m \leftarrow [\Amat\xvec^{i} + \bvecu^{i-1}]_m$.
\begin{equation}
\hspace{0.4in}u_m^{i} \leftarrow \argmin_u f(\phifun{u}{s_m}{y_m}) + \tfrac{\muu}{2}|u-d_m|^2.\label{eq:admmuupdate}
\end{equation}
\EndFor
\begin{equation}
\bvecu^{i} \leftarrow \bvecu^{i-1} + (\Amat\xvec^{i} - \uvec^{i}).\hspace{1.25in}\label{eq:admmbupdate}
\end{equation}
\If{$\|\xvec^{i} - \xvec^{i-1}\| < \epsilon_\ADMM$}
\Break
\EndIf
\EndFor
\end{algorithmic}
\caption{ADMM method for solving~\eqref{eq:PhixsAL}.}
\label{alg:ADMM}
\end{algorithmDSW}
We initialize ADMM 
using the last $\xvec$ 
from the previous iteration 
of the majorize-minimize algorithm. 
Then, $\uvec^0 \leftarrow \Amat\xvec^0$, 
leaving $\bvecu^0 = \zerovec$.
Methods 
for updating $\xvec$ 
and $\uvec$ 
depend on the specific $\Amat$ 
and $f(\cdot)$ used. 
We provide details 
for the range of cases 
explored in this paper.

\subsection{Updating $x$}\label{ssec:xupdate}


The update for $\xvec$ 
in the preceding ADMM framework
has the standard synthesis form 
of compressed sensing (CS) 
that has been extensively studied
previously~\cite{chen:98:adb,donoho:03:osr,tropp:04:gig}. 
Various CS algorithms 
may be appropriate, 
depending on $\Amat$'s structure.

If $\Amat$ is left-unitary, 
so $\Amat'\Amat = \Imat$, 
the least-squares term 
in~\eqref{eq:admmxupdate} simplifies 
to $\|\xvec - \Amat'(\uvec^i - \bvecu^i)\|_2^2$, 
plus a constant term 
(zero when $\Amat$ is 
also right-unitary), 
and updating $\xvec$ becomes 
soft thresholding:
$\xvec_n^{i+1} \leftarrow \soft{[\Amat'(\uvec^i - \bvecu^i)]_n}{\tfrac{\beta}{\muu}}$, 
where
\begin{equation}
\soft{x}{\tau} = \tfrac{x}{|x|}\max\{|x|-\tau, 0\}.
\label{eq:softthresh}
\end{equation}
When $\Amat$ is not left-unitary, 
an iterative algorithm 
like FISTA~\cite{beck:09:afi} 
may be nested 
within the ADMM method.
Algorithm~\ref{alg:FISTA} 
describes the FISTA implementation 
that approximately solves~\eqref{eq:admmxupdate}.
\begin{algorithmDSW}[!tbp]
\begin{algorithmic}
\Require $I_\FISTA$, $\xvec^0$, $\uvec$, $\bvecu$, $\beta$, $\muu$.
\State $\zvec^0 \leftarrow \xvec^0$, 
$t^0 \leftarrow 1$, 
and compute $c$ 
such that $c\Imat \succeq \muu\Amat'\Amat$.
\For{$i = 1:I_\FISTA$}
\begin{align}
\xvec^{i} &\leftarrow \soft{\zvec^{i-1} + \tfrac{\muu}{c}\Amat'(\uvec-\bvecu-\Amat\zvec^{i-1})}{\tfrac{\beta}{c}}.\hspace{0.15in}\label{eq:fistaxupdate}\\
t^{i} &\leftarrow (1+\sqrt{1+4(t^{i-1})^2})/2.\label{eq:fistatupdate}\\
\zvec^{i} &\leftarrow \xvec^{i} + \tfrac{t^{i-1}-1}{t^{i}}(\xvec^{i}-\xvec^{i-1}).\label{step:fistazupdate}
\end{align}
\EndFor
\end{algorithmic}
\caption{FISTA implementation for solving~\eqref{eq:admmxupdate}.}
\label{alg:FISTA}
\end{algorithmDSW}
When $\Amat$ is left or right unitary, 
$c = \muu$.
In other cases, $c$ is the maximum singular value of $\Amat$ 
and may be precomputed 
using power iterations.

This framework can be extended 
to analysis-form sparsity 
and other additively separable regularizers
by replacing the penalty $\|\xvec\|_1$ 
in the original objective~\eqref{eq:Psix}, 
the majorizer~\eqref{eq:Phixs}, 
and the augmented Lagrangian~\eqref{eq:PhixsAL} 
with the prior $R(\Gmat\xvec) = \sum_i r([\Gmat\xvec]_i)$, 
where $r(\cdot)$ is a potential function, 
and $\Gmat$ is an analysis transform.
The $\xvec$-update step for ADMM becomes
\begin{equation}
\xvec^{i+1} \leftarrow \argmin_\xvec \beta R(\Gmat\xvec) + \tfrac{\muu}{2}\|\Amat\xvec-(\uvec^i-\bvecu^i)\|_2^2.
\label{eq:admmxupdateanalysis}
\end{equation}
When $\Gmat$ is square and invertible, 
and the inverse $\Gmat^{-1}$ 
is readily available,
synthesis-form techniques apply.
Otherwise, 
one may nest 
within the ADMM framework 
almost any of the well-studied methods 
from the literature 
such as split Bregman iteration~\cite{goldstein:09:tsb} 
or analysis-form extensions 
of iterative methods 
like MFISTA~\cite{beck:09:fgb,wang:08:ana,tan:14:sad}.
When the proximal operator 
for $r(\cdot)$ does not have a closed form, 
proximal algorithms may also be used~\cite{parikh:13:pa}.
Alternatively, 
one may ``smooth'' 
a nonsmooth regularizer 
(using corner rounding), 
and apply gradient-based methods 
like nonlinear conjugate gradients~\cite{fletcher:64:fmb}. 

In any case, 
we can leverage 
the substantial literature 
on sparse reconstruction 
to update $\xvec$ 
within our ADMM framework. 
By using a majorizer 
and variable-splitting, 
we cast the sparse regularization component 
of the reconstruction problem 
in this well-studied form, 
without regard to the noise model
used in the data fit term 
of the original problem.

\subsection{Updating u}\label{ssec:uupdate}


An important consequence 
of the choice of variable-splitting 
is that the objective function 
for updating the auxiliary vector $\uvec$ 
is additively separable.
Thus, the update can be performed element-by-element.
Since $f(\cdot)$ is 
monotone nondecreasing, 
and $\phifun{u}{s_m}{y_m}$ 
is the pointwise maximum 
of two functions (for $y > 0$), 
we can write
$f(\phifun{u}{s_m}{y_m})$ 
as $\max\{\fpfun{u},\ \fnfun{u}\}$, 
where
\begin{align}
\fpfun{u}\ &\defeq\ \tfrac{\muu}{2}|u - d_m|^2 + f(\hpfun{u}{y_m}),
\label{eq:fpfun}\\
\fnfun{u}\ &\defeq\ \tfrac{\muu}{2}|u - d_m|^2\nonumber\\
&\quad + \begin{cases} 0, & y \leq 0,\\
f(\phinfun{u}{s_m}{y_m}), & 0 \leq |s|^q < y,\\
f(\phinfun{u}{\sbar_m}{y_m}), & 0 < y \leq |s|^q,\end{cases}\label{eq:fnfun}
\end{align}
and $d_m = [\Amat\xvec+\bvecu]_m$.
Updating $u_m$ is equivalent 
to solving 
\begin{equation}
\argmin_{u,T} T, \subjto \fpfun{u} \leq T, \fnfun{u} \leq T.
\label{eq:uupdatecon}
\end{equation}
The minimizing $T$ corresponds 
to the function value 
of $f(\phifun{u}{s_m}{y_m})$ 
at its minimum. 
The Lagrangian 
of this constrained problem 
is $T + \gamma_+(\fpfun{u}-T) + \gamma_-(\fnfun{u}-T)$, 
with Lagrange multipliers 
$\gamma_+,\gamma_- \geq 0$. 
Differentiating yields 
$\gamma_+ + \gamma_- = 1$,
and three possibilities exist:
\begin{enumerate}
\item When $\gamma_+ = 1$, and $\gamma_- = 0$, 
$\fpfun{u} = T$, 
and $\fnfun{u} < T$, 
so the optimal $u = u_+$ 
minimizes $\fpfun{u}$ 
and satisfies $\fpfun{u_+} > \fnfun{u_+}$. 
\item When $\gamma_+ = 0$, and $\gamma_- = 1$, 
the optimal $u = u_-$ 
minimizes $\fnfun{u}$ 
and satisfies $\fnfun{u_-} > \fpfun{u_-}$.
\item When $\gamma_+,\gamma_- > 0$, 
both $\fpfun{u}$ 
and $\fnfun{u}$ 
equal $T$. 
The optimal $u = u_\pm$ 
minimizes both 
of these functions 
along the curve $\fpfun{u} = \fnfun{u}$. 
\end{enumerate} 

The optimal values of $u$ 
for each case 
are computed analytically 
for $f(\cdot)$ corresponding
to the Gaussian and Laplace distribution functions 
in~\eqref{eq:nLgaussian} 
and~\eqref{eq:nLlaplace} 
on squared-magnitude measurements 
($q = 2$). 
Dropping subscripts, 
for $p = 1$, $q = 2$,
\begin{align}
u_+ &= \tfrac{\muu}{2+\muu}d,\\
u_- &= \tfrac{2s}{\muu}+d,\ \text{and}\\
u_\pm &= \sqrt{2(y+|s|^2)}\phase{((2+\muu)s + \muu d)} - s.
\end{align}
For $p = q = 2$,
\begin{align}
u_+ &= \rootop([4,0,(\muu-4y),-\muu|d|])\phase{d},\\
u_- &= (\Re{\ubar} + \imath\Im{\ubar})\phase{s},\\
u_\pm &= (c_0e^{\imath\theta} - \sbar)\phase{s},\ \text{where}\ c_0 = \sqrt{2(y+|s|^2)},\\
\nonumber\theta &= \rootop([(\tfrac{r_2}{r_1^2}\sin\alpha),(2\tfrac{r_2}{r_1^2}\cos\alpha+4),0,\\
\nonumber&\quad (2\tfrac{r_2}{r_1^2}\cos\alpha-4),-\tfrac{r_2}{r_1^2}\sin\alpha]),\\
\nonumber c_1 &= c_0^2+|s|^2-y, r_1 = 2c_0|s|,\\
\nonumber &\quad \text{and $r_2$ and $\alpha$ are
the magnitude and phase of}\\
\nonumber &\quad c_0(4c_1|s|+\muu(|s|+d\nphase{s})).
\end{align}
When calculating $u_+$ and $u_\pm$ 
for the Gaussian case, 
the root used 
is the one 
whose corresponding $u$ 
minimizes $f_+(u)$.
These expressions 
are derived in the appendices.

\section{Parameter Tuning}\label{sec:parameters}

%

The regularization parameter $\beta$ 
controls the level 
of sparsity
in the reconstructed signal. 
Additionally, 
the ADMM penalty parameter $\mu$ 
impacts the convergence rate 
of the inner ADMM algorithm, 
and thus, 
the overall algorithm.
This section explores 
the influence of these parameters.

Our simulations 
consist of generating a length-$N$ sparse signal 
with $K$ nonzero coefficients, 
$M$ measurements of the DFT of that signal, 
performing the reconstructions, 
and comparing the reconstructed signals 
against the true signal.
The sparse support of our signal 
is chosen at random, 
and the amplitude and phase 
of each of nonzero coefficient 
are randomly sampled uniformly 
between $0.5$ and $1$ (for amplitude) 
and $0$ and $2\pi$ (for phase). 
Then, $M$ noise-free measurements are 
randomly selected  
from the squared-magnitude 
of the signal's DFT coefficients. 
Randomly selected outliers are set 
to have an amplitude 
of twice the maximum measurement.
This model does not exactly match 
our Laplace noise model, 
thus avoiding an ``inverse crime.''
The reconstructions are performed 
using multiple initializations, 
and the ``best'' reconstructed signal 
for each method is retained.
For the proposed method, 
the best reconstruction yields 
the lowest value of $\Psi(\xvec)$.

Sparsity and Fourier coefficient magnitudes 
are insensitive to spatial shifts, reversal, 
and global phase,
so we find the best alignment/reversal 
and global phase 
for the reconstructed signals 
before evaluation.
The best alignment is identified 
for both the reconstructed signal 
and its reversed version 
by cross-correlation with the true signal. 
A global phase term 
is then estimated 
from the version with the best alignment.
For evaluation, 
a sparse threshold of $0.05$ is used 
to identify the sparse support 
of the reconstructed signal. 
The sparse support 
of a correctly detected signal 
matches that of the true signal.

\begin{figure}[!tbp]
	\centering
	\includegraphics[width=3.2in]{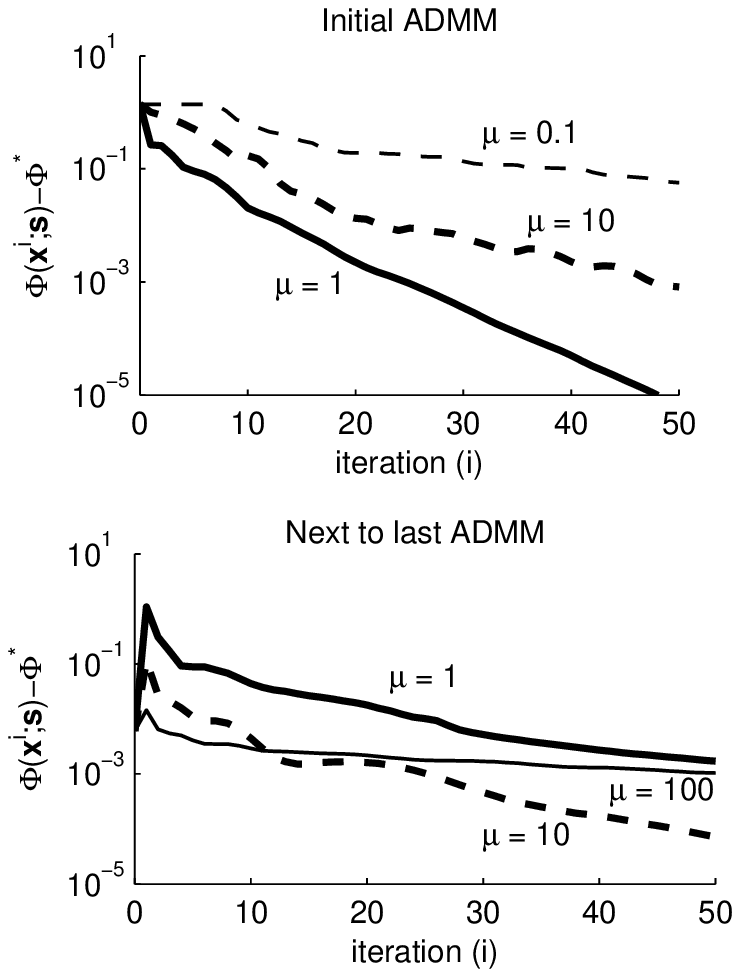}\\
	\vspace{-0.1in}
	\caption{The objective function $\Phi(\xvec^i;\svec)$, relative to converged objective value $\Phi^*$, is plotted 
		versus ADMM iteration $i$ for both the first and the next-to-last run of ADMM, 
		for the Laplace ($p = 1$) noise model.}
	\label{fig:mus_noisefree_p1q2}
\end{figure}
\begin{figure}[!tbp]
	\centering
	\includegraphics[width=3.2in]{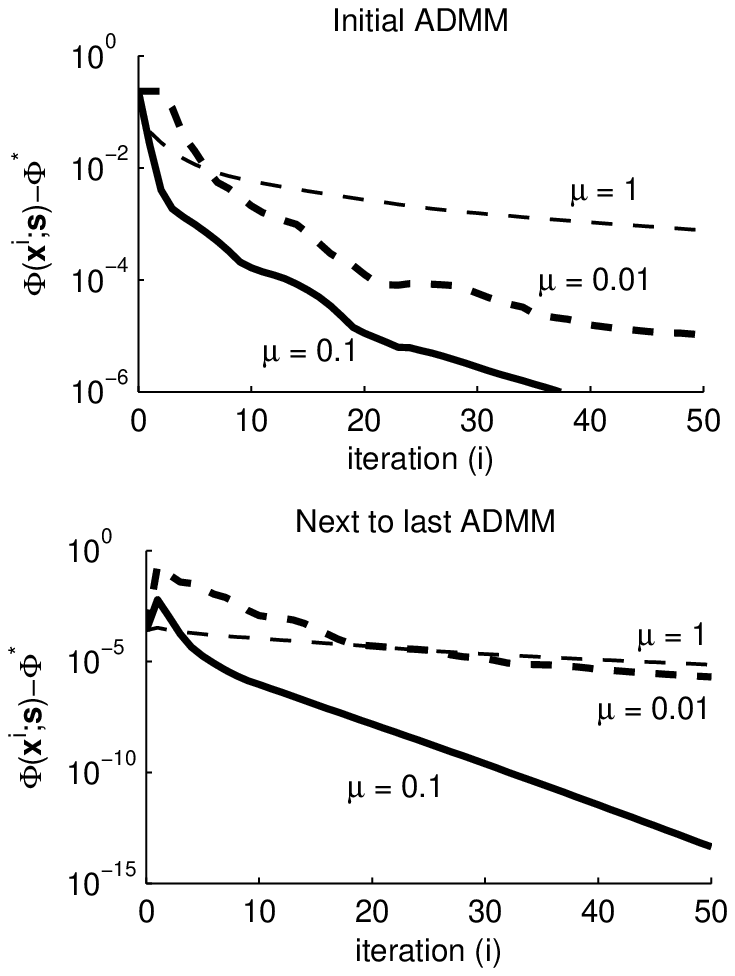}\\
	\vspace{-0.1in}
	\caption{The objective function $\Phi(\xvec^i;\svec)$, relative to converged objective value $\Phi^*$, is plotted 
		versus ADMM iteration $i$ for both the first and the next-to-last run of ADMM, 
		for the Gaussian ($p = 2$) noise model.}
	\label{fig:mus_noisefree_p2q2}
\end{figure}
\begin{figure}[!tbp]
\centering
\subfloat[][Laplace ($p = 1$) noise model.]{\includegraphics[width=3.2in]{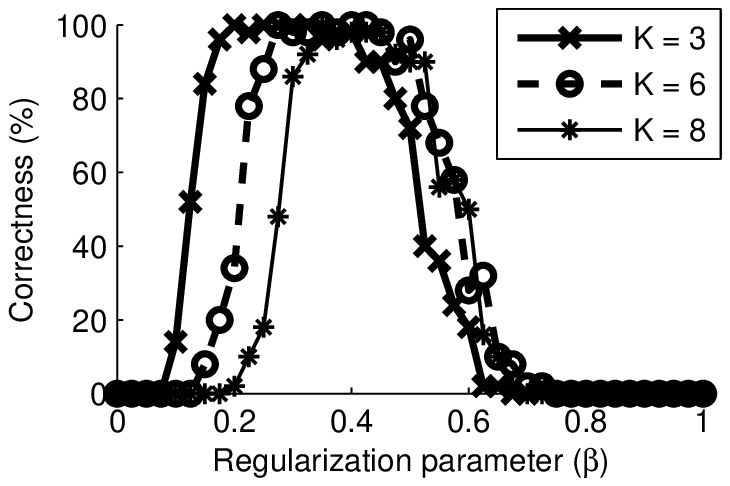}}\\
\subfloat[][Gaussian ($p = 2$) noise model.]{\includegraphics[width=3.2in]{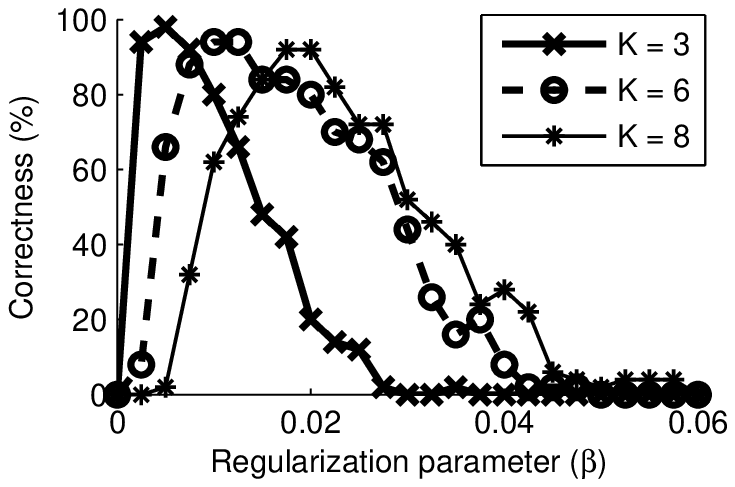}}\\
\caption{The percentage of $50$ trials reconstructed correctly 
is plotted versus regularization parameter $\beta$ 
for varying signal sparsity levels $K$, 
for the proposed reconstruction 
with (a) Laplace ($p = 1$) and (b) Gaussian ($p = 2$) noise models.}
\label{fig:betas_noisefree_proposed}
\end{figure}
In our first experiment, 
we reconstruct a 128-element 1D signal 
using both Laplace ($p = 1$)
and Gaussian noise models ($p = 2$) 
and ADMM with different values 
of penalty parameter $\mu$, 
for different degrees of sparsity 
and numbers of measurements. 
Since the optimal ADMM parameter 
may differ between earlier 
and later majorizer minimization iterations, 
we compare the convergence rates 
for different choices of $\mu$ 
in both the initial 
and next-to-last runs of ADMM.
Figures~\ref{fig:mus_noisefree_p1q2} 
and~\ref{fig:mus_noisefree_p2q2}
portray, 
for sparsity $K = 6$, $M = 64$ noiseless measurements,
and Laplace ($p = 1$) and Gaussian ($p = 2$) noise models, 
respectively,
the objective function convergence rates 
over $I_\text{ADMM} = 50$ ADMM iterations
for the three best choices of $\mu$, 
relative to the best objective 
function value observed over $200$ ADMM iterations.
Running the same experiment for different sparsity $K = 8$ 
and $M = 128$ measurements yield similar results 
to the example shown, 
with the same optimal $\mu$'s.
In this experiment, we observe the optimal choice of $\mu$ 
for the proposed method with $p = 1$ 
does not change much from the initial to the next-to-last run of ADMM, 
changing only from $\mu = 1$ 
to $\mu = 10$. 
However, a minor change in $\mu$ can make a huge difference 
in convergence rate, 
especially in later iterations, 
so using an adaptive scheme 
like the heuristic method 
described in~\cite{boyd:10:doa} 
would help maintain fast convergence.
The optimal choice of $\mu$ appears more stable 
in the $p = 2$ case, 
as $\mu = 0.1$ yields the fastest objective function convergence 
for both early and later runs.
The optimal choices of $\mu$ 
in both instances do not appear to change 
with sparsity $K$ or number of measurements $M$, 
or the associated changes in $\beta$, 
so we used these values of $\mu$ 
throughout the experiments that follow.

\begin{figure}[!tbp]
\centering
\includegraphics[width=3.2in]{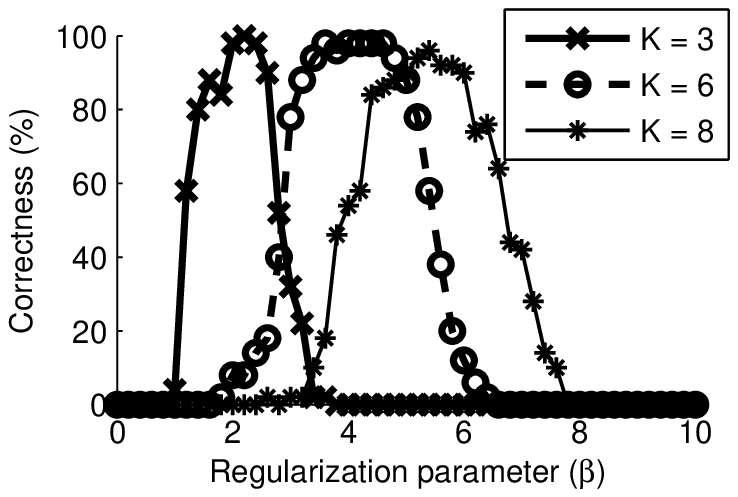}\\
\caption{The percentage of $50$ trials reconstructed correctly 
is plotted for the modified sparse Fienup (L$_1$-Fienup) method 
that projects the image domain reconstruction onto the $1$-norm ball $\|\xvec\|_1 \leq \beta$, versus constraint parameter $\beta$ 
for varying signal sparsity levels $K$.}
\label{fig:betas_noisefree_sf}
\end{figure}
\begin{figure*}[!tbp]
\centering
\includegraphics[width=6.5in]{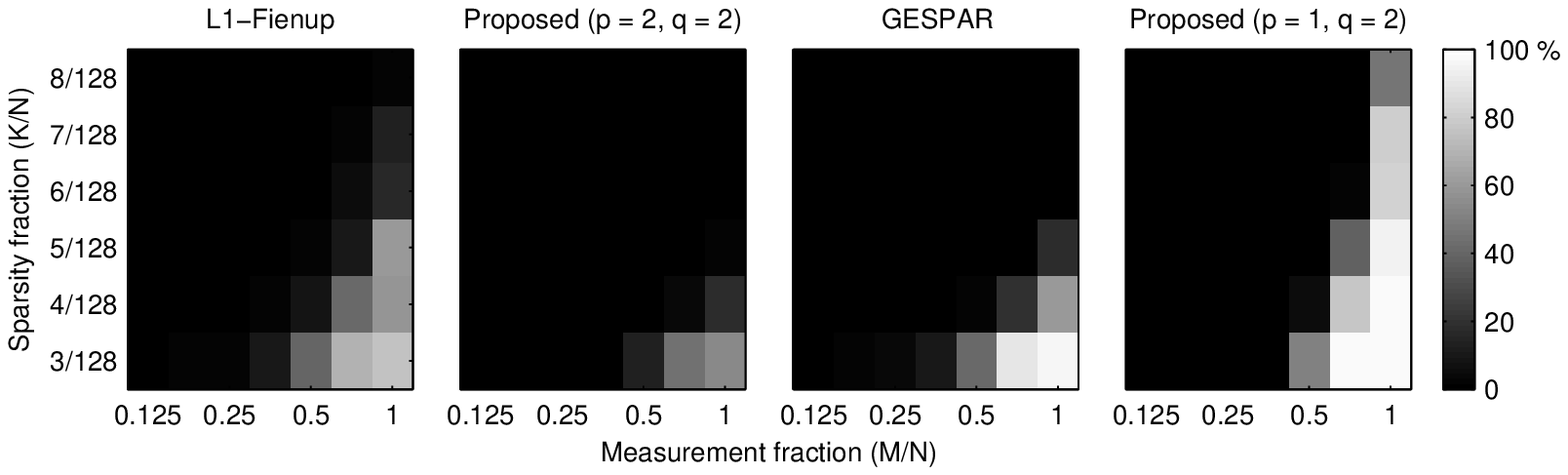}\\
\caption{The percentage of $50$ trials reconstructed correctly 
is given for the modified sparse Fienup (L$_1$-Fienup) method, 
GESPAR, 
and the proposed method with both Gaussian $p = 2$ 
and Laplace $p = 1$ noise models, 
for a range of  measurement and sparsity fractions.}
\label{fig:MC_5out_correct}
\end{figure*}
To study the effects 
of varying $\beta$ 
on the performance 
of the algorithm, 
we focus on reconstructing 1D signals 
using either Laplace 
or Gaussian noise models 
for varying degrees of sparsity.
Here, we used $40$ random initializations 
for both $p = 1$ and $p = 2$ cases.
In~\cite{weller:14:pro}, 
the optimal range of $\beta$ 
for the proposed method 
with $p = 1$, $q = 2$ 
is shown to scale roughly linearly 
with the number of measurements. 
Here, we evaluate the proposed method 
with both Laplace and Gaussian noise models 
for $M = N = 128$ noise-free measurements.
Figure~\ref{fig:betas_noisefree_proposed} plots 
the percentage of $50$ trials reconstructed 
with the correct support 
versus the regularization parameter 
for different sparsity levels $K = 3, 6, 8$, 
for both noise models.

For comparison, 
we also evaluate the sparse Fienup method, 
with the image-domain projection modified 
to project the signal onto the $1$-norm ball with $\|\xvec\|_1 \leq \beta$, 
for different values of regularization parameter $\beta$ 
in Fig.~\ref{fig:betas_noisefree_sf}.
This modification replaces the hard-thresholding sparse projection 
onto the $0$-``norm'' ball 
with a $1$-norm projection more closely aligned 
with the sparsity penalty used in the proposed method.
We call this modified method L$_1$-Fienup 
in the results that follow.

This L$_1$-Fienup method 
exemplifies the great importance 
the choice of $\beta$ has 
on the reconstruction quality. 
Not only does $\beta$ greatly influence 
the chance of correct support detection, 
but the optimal choice of $\beta$ 
greatly depends on the sparsity $K$ of the signal. 
The optimal $\beta$ for $K = 8$ would work extremely poorly 
for $K = 3$, 
and vice versa.
The dependence on $\beta$ 
of the proposed method 
is very similar, 
for both noise models. 
The $p = 1$ case 
demonstrates less variation 
in the correctness 
as a function of $\beta$ 
than the $p = 2$ case, 
but a reasonably good choice of $\beta$ 
is necessary for correct reconstruction 
with either noise model.
The optimal choices of $\beta$ were computed 
for all the values of $K$, 
without noise, 
used in the experiments that follow, 
including the 2D image comparisons.

\section{Monte Carlo Comparisons (1D)}\label{sec:mccomp}


We compared phase retrieval methods 
using Monte Carlo simulations 
for different values of sparsity $K$ 
and number of measurements $M$, 
with $50$ trials each.
We compare the proposed method 
with both $p = 1$ (Laplace) and $p = 2$ (Gaussian) data fit models 
against both the L$_1$-Fienup method 
described previously 
and the GESPAR greedy method 
recently developed for the Gaussian noise model. 
Table~\ref{table:methods} highlights 
the differences between the four methods.

\begin{table}[!tbp]
\centering
\caption{Comparison of Reconstruction Methods}
\label{table:methods}
\begin{tabular}{lccc}
Method & Implementation & Sparsity & Noise Model \\\hline
L$_1$-Fienup & alternating & $1$-norm & Gaussian\\
& projections & & \\\hline
Proposed ($p = 2$) & MM, ADMM & $1$-norm & Gaussian\\\hline
GESPAR & greedy & $0$-``norm'' & Gaussian\\\hline
Proposed ($p = 1$) & MM, ADMM & $1$-norm & Laplace\\\hline
\end{tabular}
\end{table}

\begin{figure*}[!tbp]
\centering
\includegraphics[width=6.5in]{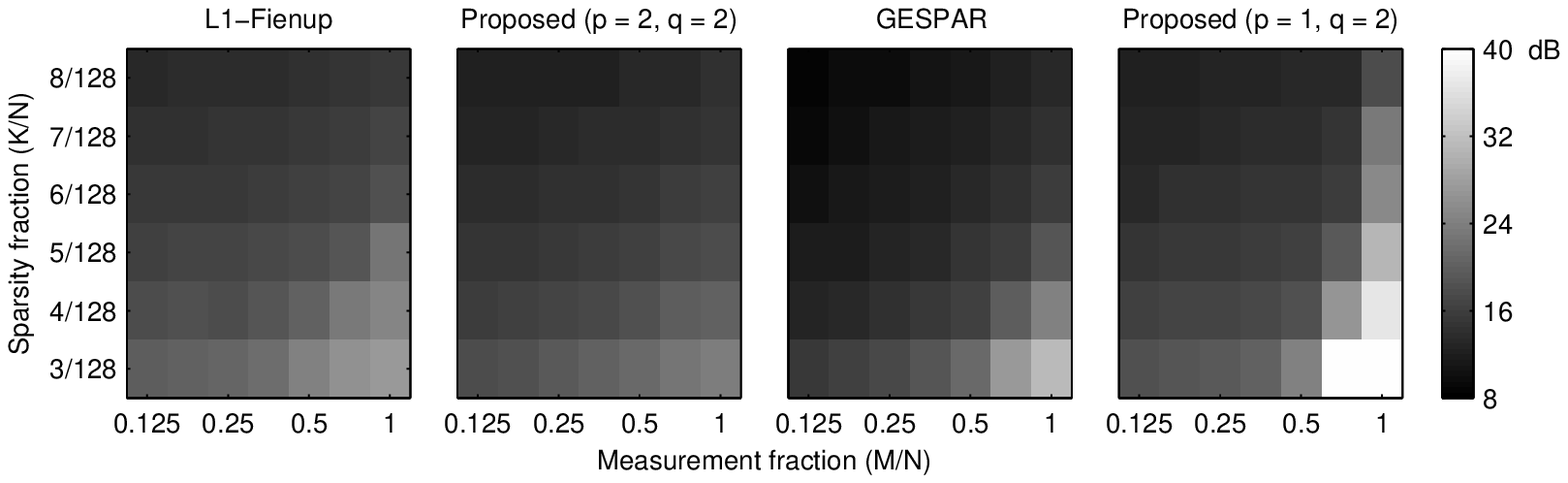}\\
\caption{The average PSNR, in dB, over $50$ trials 
is given for the modified sparse Fienup (L$_1$-Fienup) method, 
GESPAR, 
and the proposed method with both Gaussian $p = 2$ 
and Laplace $p = 1$ noise models, 
for a range of  measurement and sparsity fractions.}
\label{fig:MC_5out_psnr}
\end{figure*}

These methods all use multiple initializations, 
with $40$ initializations for L$_1$-Fienup 
and the proposed method with $p = 1$, 
and with $50$ initializations 
for the proposed method with $p = 2$. 
The GESPAR method tests different initializations 
until the sparse signal achieves data discrepancy 
below a fixed threshold.
The percentage of trials with correctly reconstructed 
(detected) signal supports 
is shown for all four methods 
in Fig.~\ref{fig:MC_5out_correct},
as a function of both number of measurements $M$ 
(with five outliers)
and sparsity of the true signal $K$. 
In addition, 
the average mean squared error (MSE)
is reported in Fig.~\ref{fig:MC_5out_psnr} 
in terms of peak signal to noise ratio, 
$\text{PSNR} = 10\log_{10} \tfrac{1}{\text{MSE}}$, 
where the maximum true signal amplitude is one.
To achieve the results shown, 
we had to increase the support detection threshold 
to $0.2$ for the proposed method with $p = 2$ only, 
suggesting inadequate convergence for the Gaussian model.

\begin{figure*}[!tbp]
\centering
\subfloat[][Percentage of correct detections for $K = 3$ sparsity (out of $N = 128$)]{\includegraphics[width=6.5in]{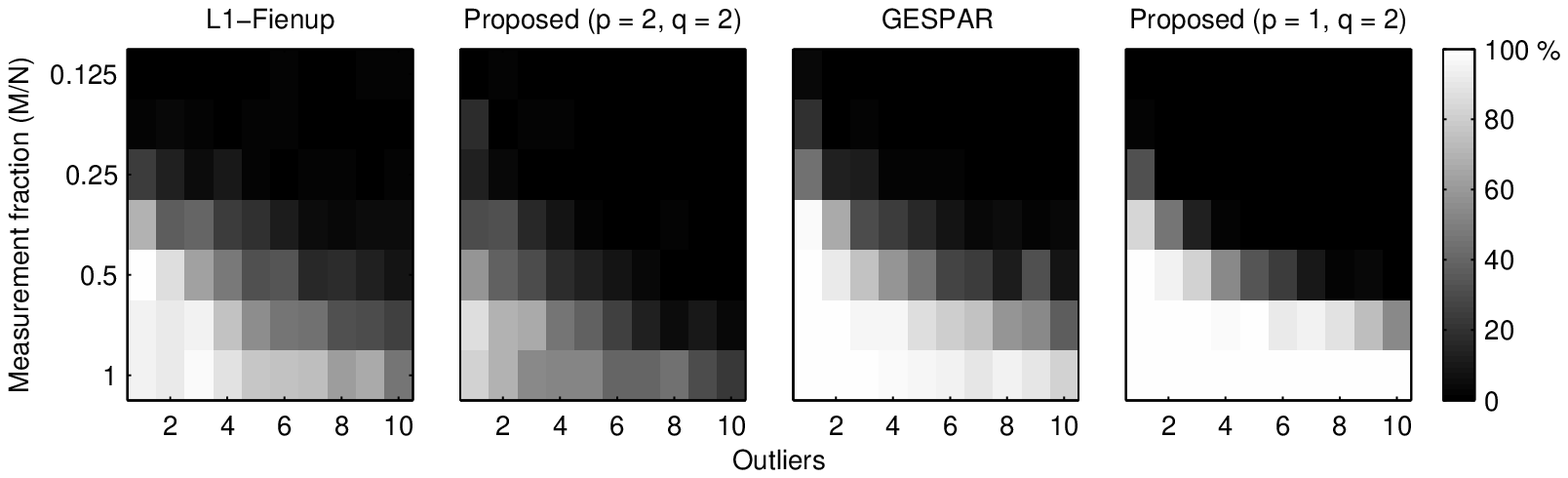}}\\
\subfloat[][PSNRs (in dB) for $K = 3$ sparsity (out of $N = 128$)]{\includegraphics[width=6.5in]{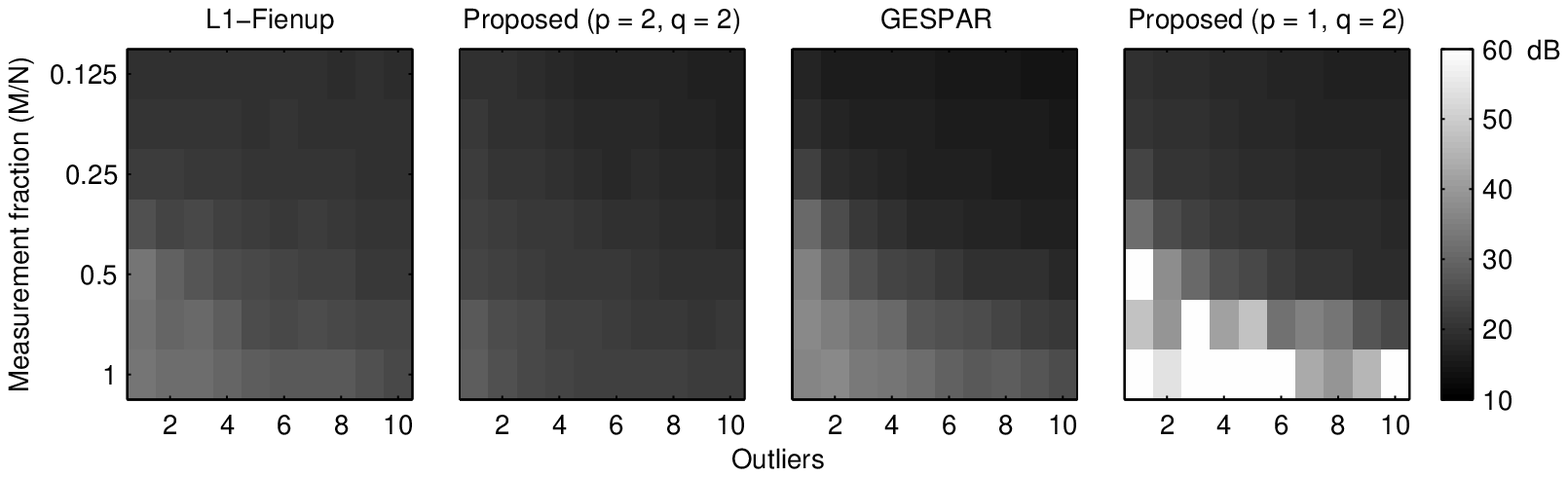}}\\
\subfloat[][Percentage of correct detections for $K = 5$ sparsity (out of $N = 128$)]{\includegraphics[width=6.5in]{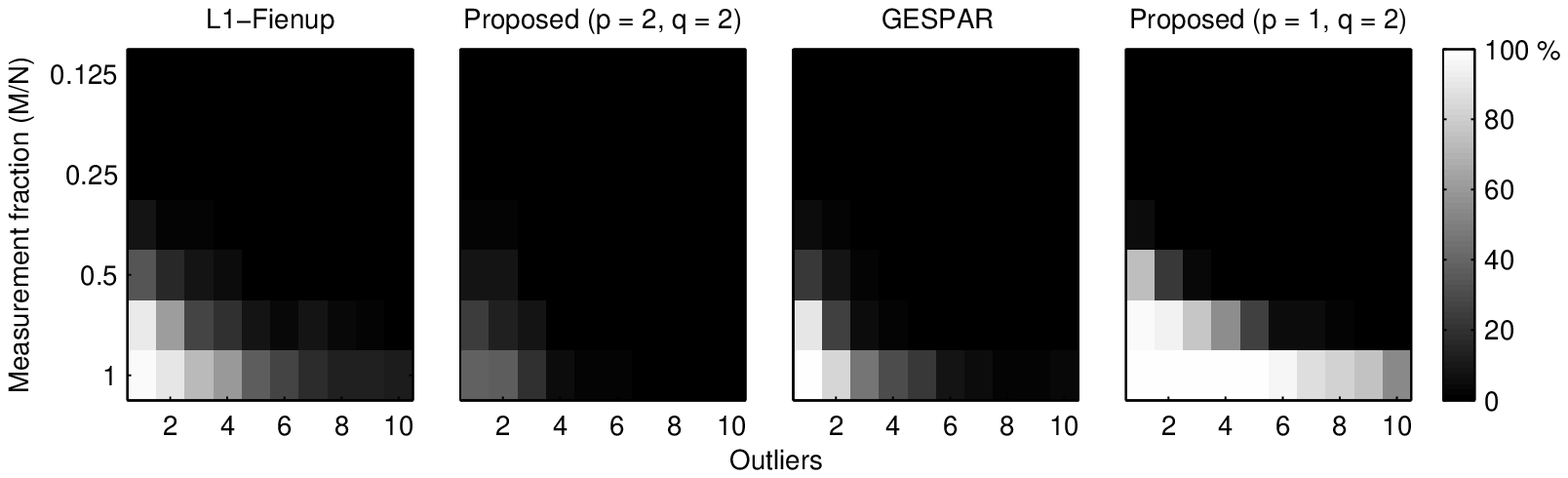}}\\
\subfloat[][PSNRs (in dB) for $K = 5$ sparsity (out of $N = 128$)]{\includegraphics[width=6.5in]{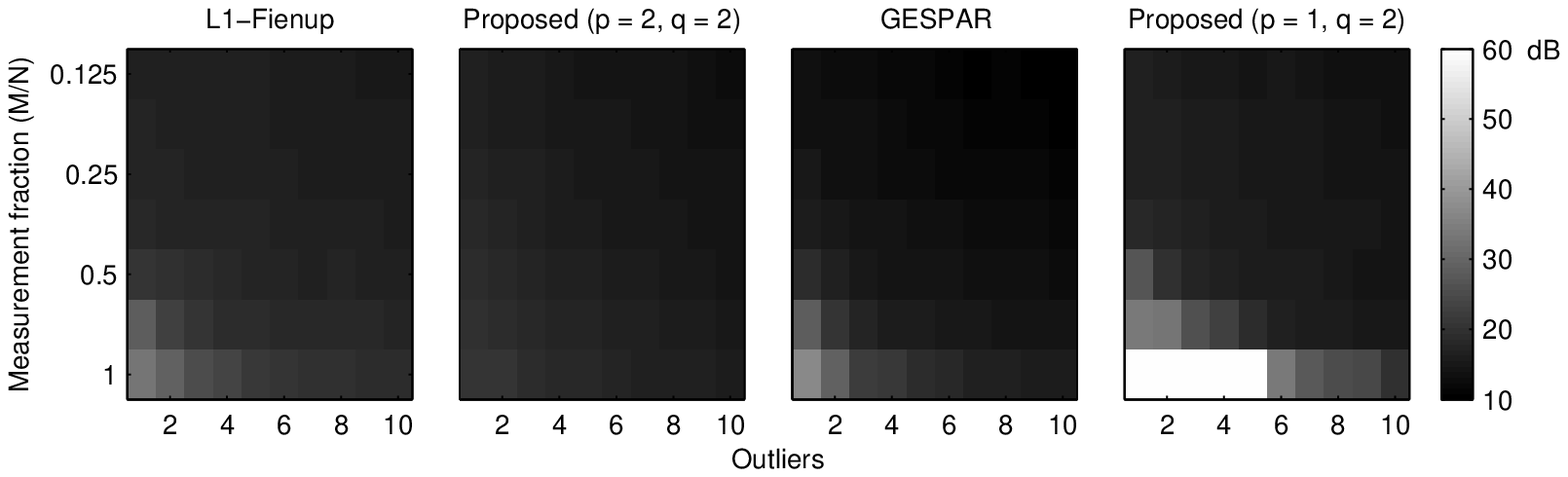}}\\
\caption{The correctness and average PSNR, in dB, are plotted for $50$ trials of the L$_1$-Fienup, GESPAR, and proposed algorithm with both models, for between $1$ and $10$ outliers out of $M = 16$ to $M = 128$ measurements, for $N=128$-length signals with sparsities $K = 3$ (top) and $K = 5$ (bottom).}
\label{fig:MC_outliers}
\end{figure*}

These results suggest that 
the proposed method with the Laplace model $p = 1$, 
which more closely models 
the outliers in the measurements, 
attains the best performance 
of the four methods tested, 
in terms of both support recovery 
and PSNR.
Figure~\ref{fig:MC_outliers} depicts 
trends in the correctness and PSNRs 
of the four methods 
as the number of outliers increases.

\section{Image Comparisons (2D)}\label{sec:imcomp}


\begin{figure}[!tbp]
\centering
\includegraphics[width=3.2in]{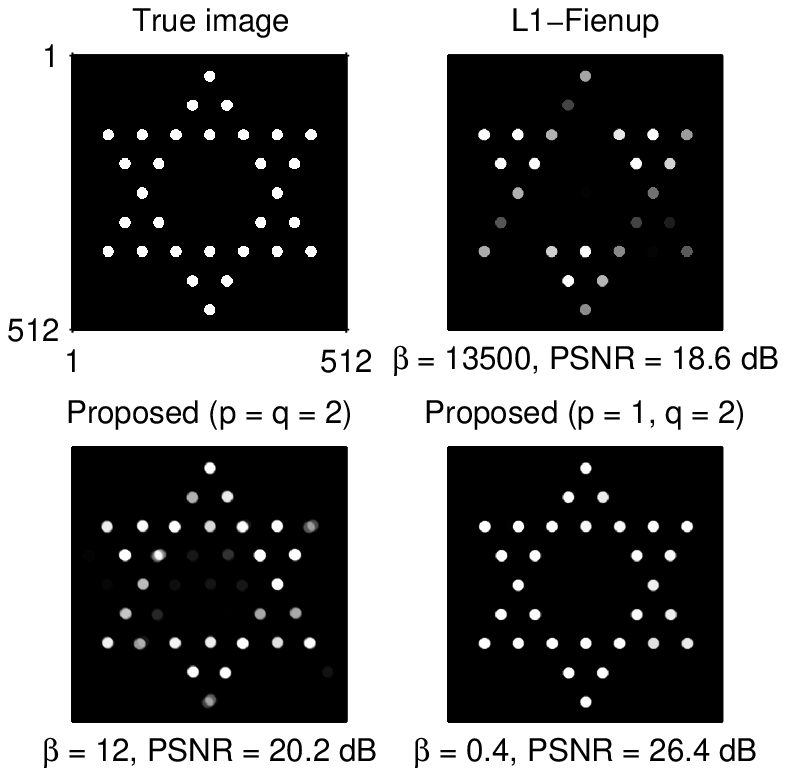}\\
\caption{The best reconstructions (as a function of regularization parameter $\beta$)
for L$_1$-Fienup and the proposed method 
with both noise models 
are shown 
for the $512\times 512$-pixel star of David phantom, 
from $M = N/2$ measurements, 
with $10$ outliers.}
\label{fig:SOD_10out}
\end{figure}

To demonstrate 
how the proposed method performs 
for image reconstruction, 
we examine the two-dimensional case 
with undersampled measurements 
corrupted by $10$ outliers. 
The $512\times 512$-pixel star of David phantom 
used in~\cite{weller:14:pro} 
is inspired by the real example image 
shown in~\cite{szameit:12:sbs}.
The DFT of this image 
is randomly undersampled by a factor of two 
and reconstructed using both the proposed 
and the state-of-the-art algorithms.
The reconstruction using the proposed method 
with the Laplace model produces a nearly-perfect image.
The L$_1$-Fienup method yields an image 
with degraded or missing dots, 
especially in the lower left and right triangles, 
and near the top. 
The proposed method with a Gaussian model 
produces a more consistent reconstruction 
than the L$_1$-Fienup method, 
but a number of additional dots near the center 
are visible. 
The $p=2$ case shown uses $\mu = 1$; 
setting $\mu=0.1$ degrades reconstruction quality
in this case.

\section{Discussion}\label{sec:discussion}

Phase retrieval relies heavily 
on side information 
to reproduce a quality image. 
We employ sparsity in the image domain, 
or dictionary-based sparsity, 
to identify the best image 
among all those that share 
the same magnitude Fourier spectrum. 
Resolving this ambiguity becomes even more challenging 
in the face of measurement noise, 
especially outliers, 
and undersampling the Fourier spectrum.
The proposed method
using a Laplace noise model 
excels at reconstructing images 
despite these conditions, 
greatly improving upon other techniques 
and the Gaussian noise model 
for such data.

Parameter tuning does not appear to be more challenging 
than with existing methods 
in our simulations, 
especially considering the actual sparsity $K$ usually is not known.
Future research concerning automatic calibration 
and generalization of parameter selection is ongoing~\cite{weller:14:mcs}, 
and phase retrieval would appear to be an excellent application, 
based on its sensitivity to the choice of regularization parameter $\beta$.
Additionally, using an adaptive heuristic 
for the penalty parameter $\mu$ appears 
to offer satisfactory convergence 
without substantial additional tuning.
Further experiments on larger sets 
of different data are necessary, however, 
to draw more general conclusions 
about these parameters.

Paired with parameter selection, 
multiple initializations are also essential 
to overcome the nonconvexity 
of the inverse problem 
and find a reasonable (hopefully) global solution. 
Although we investigated promising techniques 
for initializing our method, 
like Wirtinger flow~\cite{candes:14:prv},
randomly selecting multiple initial majorization vectors $\svec^0$ 
appears to be more robust.
However, using multiple initial choices 
for $\svec^0$ 
proportionally increases 
in computational burden. 
Combined with the multi-layered nature 
of the proposed algorithm,
the overall reconstruction time becomes an issue 
in higher dimensions. 
In the 2D image reconstruction case, 
running a reconstruction for a single choice of $\beta$
(and multiple choices were used 
for parameter tuning reasons) 
consumed several hours on a modern processor running MATLAB.
Efforts to accelerate convergence 
of the proposed algorithm, 
such as applying momentum~\cite{nesterov:83:amo}, 
would be well worth further study.

Computational costs aside, 
our method clearly outperforms 
the L$_1$-modified Fienup method 
and GESPAR, 
when outliers are present in the data. 
Our method improves both the likelihood of correct support recovery 
and the overall normalized MSE 
in both 1D Monte Carlo 
and 2D image simulations.

Our framework may extend 
to more general regularizers $R(\Gmat\xvec)$ 
via generalization 
of the $\xvec$-update step 
to nest an algorithm like split Bregman iteration.
Such a modification would enable analysis-form sparsity regularization 
with total variation 
or undecimated wavelets.
The $\xvec$-update step 
also can accommodate other priors or constraints, 
like support information or nonnegativity, 
by using an appropriate nested algorithm 
in place of soft-thresholding or FISTA.

\section{Conclusion}\label{sec:conclusion}

The key contributions of this paper are two-fold. 
First, a general framework was proposed 
that extends phase retrieval reconstruction 
to measurements corrupted by outliers in the data. 
A multi-layered implementation 
of this general framework
was developed 
featuring multiple initializations, 
majorization-minimization, 
and ADMM.
Secondly, the sensitivities to both the regularization 
and penalty parameters present 
in the reconstruction framework and algorithms 
were studied, 
aiming to provide a fast, robust, and correct reconstruction method.
The analysis of the proposed method then shifted 
to a direct comparison against competing methods 
including an L$_1$-modified sparse Fienup method 
and the greedy algorithm known as GESPAR.
These comparisons included both 
a 1D Monte Carlo experiment 
to establish quantitative 
advantages over existing methods, 
and a 2D image reconstruction 
visually demonstrating the improvements achievable 
using this method, 
even with relatively few outliers 
in the data.

\appendices

\section{Updating $u$: Squared-magnitude Laplace case}\label{app:laplace}

In this case, $f(\cdot) = (\cdot)$, and $q = 2$. 
When $y_m < 0$, 
$\fpfun{u}$ is always greater 
than $\fnfun{u}$, 
so the solution 
is always the minimizer 
of $\fpfun{u}$. 
Otherwise, 
we must consider all three cases.

Let $d = [\Amat\xvec^{i+1} + \bvecu^i]_m$, 
$s$ represent the appropriate 
choice of $s_m$ or $\sbar_m$, 
$\eta\ \defeq\ \muu/2$, 
and drop the subscripts.
Writing out $\fpfun{u}$ and $\fnfun{u}$, 
\begin{align*}
\fpfun{u} &= \eta|u - d|^2 + |u|^2 - y,\\
\fnfun{u} &= \eta|u - d|^2 + y + |s|^2 - 2|s|\Re{u\nphase{s}}.
\end{align*}

The function $\fpfun{u}$ is quadratic 
in $u$, 
so completing the square yields 
\[
\fpfun{u} = (1+\eta)|u - \tfrac{\eta}{1+\eta}d|^2 + (\tfrac{\eta}{1+\eta}|d|^2-y).
\]
Thus, $\fpfun{u}$ is minimized 
by $u_+ = \tfrac{\eta}{1+\eta}d$.

The function $\fnfun{u}$ is also a quadratic,
so
\[
\fnfun{u} = \eta|u-e|^2 + (y+|s|^2+\eta|d|^2-\eta|e|^2),
\]
where $e\ \defeq\ \tfrac{s}{\eta}+d$. 
The minimizer is simply $u_- = e$.

The minimization 
of $\fpfun{u}$ 
or $\fnfun{u}$ 
along the curve 
on which both functions 
are equal-valued, 
involves parameterizing 
this curve 
and minimizing $\fpfun{u}$ 
as a function of this parameter.
These functions are equal 
when $|u|^2-y = y+|s|^2-2|s|\Re{u\nphase{s}}$, 
which corresponds to the circle 
$|u+s|^2 = 2(y+|s|^2)$.
The parameterization then 
correponds to the angle along the circle; 
call it $\theta$. 
The curve of interest is 
$(u+s) = \sqrt{2(y+|s|^2)}e^{\imath\theta}$.
Incorporating this parameterization 
into $\fpfun{u}$ 
yields
\[\begin{split}
\fpfun{u(\theta)} &= -2\sqrt{2(y+|s|^2)}\Re{((1+\eta)s+\eta d)e^{-\imath\theta}}\\ 
&\quad + \text{constants},\end{split}
\]
which is minimized when $\theta = \angle ((1+\eta)s + \eta d)$.
So, $u_\pm = \sqrt{2(y+|s|^2)}\phase{((1+\eta)s + \eta d)} - s$.

\section{Updating $u$: Squared-magnitude Gaussian case}\label{app:gaussian}

In this case, $f(\cdot) = (\cdot)^2$, and $q = 2$.
Again, 
as with the Laplace distribution, 
when $y_m < 0$, 
$\fpfun{u} > \fnfun{u}$, 
so we always minimize $\fpfun{u}$.
Otherwise, 
we consider all three cases.

Again, 
let $d = [\Amat\xvec^{i+1} + \bvecu^i]_m$, 
$s$ represent the appropriate 
choice of $s_m$ or $\sbar_m$, 
$\eta\ \defeq\ \muu/2$, 
and drop the subscripts.
Writing out $\fpfun{u}$ and $\fnfun{u}$, 
\begin{align*}
\fpfun{u} &= \eta|u - d|^2 + (|u|^2 - y)^2,\\
\fnfun{u} &= \eta|u - d|^2 + (y + |s|^2 - 2|s|\Re{u\nphase{s}})^2.
\end{align*}

Writing $\fpfun{u}$ 
in terms 
of the magnitude $|u|$ 
and phase $\angle u$ 
of $u$, 
\[\begin{split}
\fpfun{u} &= \eta|u|^2 + \eta|d|^2 - 2\eta|u||d|\cos(\angle u - \angle d)\\
&\quad + |u|^4 - 2y|u|^2 + y^2,\end{split}
\]
which is clearly minimized 
when $\angle u = \angle d$, 
when $\cos(\angle u - \angle d) = 1$.
Then, 
$\fpfun{u}$ 
becomes a quartic equation in $|u|$, 
which has the derivative
\[
\frac{d\fpfun{u}}{d|u|} = 4|u|^3 + (2\eta-4y)|u| - 2\eta|d|.
\]
The function $\fpfun{u}$ 
is minimized either 
when the derivative is zero 
or when $|u| = 0$.
The depressed cubic equation 
will have between zero 
and three nonnegative real roots, 
which can be found analytically.
Note that if there are 
three positive real roots, 
since the cubic must be increasing 
below the least positive root, 
the derivative at $|u| = 0$ 
is negative, 
and the fourth candidate point $|u| = 0$ 
cannot be the global minimum.
The minimizer $u_+$ 
is the candidate point 
with minimum function value $\fpfun{|u|}$, 
multiplied by $\phase{d}$.

Finding a minimum 
of $\fnfun{u}$ 
is straightforward. 
Define $\ubar = u\nphase{s}$, 
and $\dbar = d\nphase{s}$. 
Then, 
\[
\fnfun{\ubar} = \eta|\ubar - \dbar|^2 + (y + |s|^2 - 2|s|\Re{\ubar})^2.
\]
Separating the real 
and imaginary parts, 
we observe
\[\begin{split}
\fnfun{\ubar} &= \eta(\Re{\ubar}-\Re{\dbar})^2 + \eta(\Im{\ubar}-\Im{\dbar})^2\\
&\quad + (y + |s|^2 - 2|s|\Re{\ubar})^2,\end{split}
\]
which is clearly minimized 
when $\Im{\ubar} = \Im{\dbar}$.
The real component 
is quadratic in $\Re{\ubar}$, 
so differentiating 
with respect to $\Re{\ubar}$
yields 
\[\begin{split}
\frac{d\fpfun{\ubar}}{d\Re{\ubar}} &= 2\eta(\Re{\ubar} - \Re{\dbar})\\
&\quad + 4|s|(2|s|\Re{\ubar} - (y+|s|^2)),\end{split}
\]
which is minimized 
at 
\[
\Re{\ubar} = \frac{\eta\Re{\dbar} + 2|s|(y+|s|^2)}{\eta + 4|s|^2}.
\]
This closed form solution 
yields
\[
u_- = (\Re{\ubar} + \imath\Im{\ubar})\phase{s}.
\]

Minimizing $\fpfun{u}$ 
along the curve $\fpfun{u} = \fnfun{u}$ 
requires parameterizing the curve.
Again, define 
$\ubar = u\nphase{s}$, 
$\dbar = d\nphase{s}$, 
and $\sbar = |s|$. 
Note that 
$\phinfun{\ubar}{\sbar}{y} = |\sbar-\ubar|^2 + (y - |\ubar|^2)$, 
where the latter term equals $B\ \defeq\ -\hpfun{\ubar}{y}$. 
Along the curve $\fpfun{\ubar} = \fnfun{\ubar}$, 
$B^2 = (B+|\sbar-\ubar|^2)^2$, 
which is true when $s = \ubar$, 
or when $|\sbar-\ubar|^2 = -2B = 2(|\ubar|^2 - y)$. 
For this second case 
to yield a nontrivial solution 
requires $B < 0$, 
which corresponds to $|\ubar|^2 > y$. 

Rearranging terms 
yields our familiar circle 
$|\ubar+\sbar|^2 = 2(y + \sbar^2)$ 
from the Laplace distribution case.
Plugging our angular parameterization 
$\ubar = c_0e^{\imath\theta} - s$, 
where $c_0 = \sqrt{2(y+\sbar^2)}$, 
into $\fpfun{\ubar}$ 
yields 
\[\begin{split}
\fpfun{\ubar(\theta)} &= (|c_0e^{\imath\theta}-\sbar|^2-y)^2 + \eta|c_0e^{\imath\theta}-\sbar-\dbar|^2\\
&= (c_0^2-2c_0\Re{e^{\imath\theta}\sbar^*} + \sbar^2 - y)^2\\
&\quad + \eta(c_0^2 + |\sbar+\dbar|^2 - 2c_0\Re{e^{i\theta}(\sbar+\dbar)^*}).\end{split}
\]
Let $c_1 = c_0^2 + \sbar^2 - y$, 
and $c_2 = c_0^2 + |\sbar+\dbar|^2$, 
so 
\[\begin{split}
\fpfun{\ubar(\theta)} &= (c_1 - 2c_0\Re{e^{\imath\theta}\sbar^*})^2\\ 
&\quad + \eta(c_2 - 2c_0\Re{e^{\imath\theta}(\sbar+\dbar)^*})\\
&= (2c_0)^2\Re{e^{\imath\theta}\sbar^*}^2\\
&\quad - 2c_0\Re{e^{\imath\theta}(2c_1\sbar+\eta(\sbar+\dbar))^*} + c_1^2 + \eta c_2
.\end{split}
\]
For convenience, 
let $r_1 = 2c_0\sbar$, 
and $r_2$ and $\alpha$ 
be the magnitude and phase
of $2c_0(2c_1\sbar+\eta(\sbar+\dbar))$.
Differentiating 
with respect to $\theta$, 
\[
\frac{d\fpfun{\ubar(\theta)}}{d\theta} = r_2\sin(\theta-\alpha) - 2r_1^2\sin\theta\cos\theta.
\]
Setting the derivative 
equal to zero,
\[
\tfrac{r_2}{r_1^2}\sin(\theta-\alpha) = \sin(2\theta).
\]
Defining $\xi$ 
such that $\theta = 2\arctan\xi$, 
we have 
$\sin\theta = \sin(2\arctan\xi) = \tfrac{2\xi}{1+\xi^2}$, 
and $\cos\theta = \cos(2\arctan\xi) = \tfrac{1-\xi^2}{1+\xi^2}$.
Thus, 
\begin{align*}
\sin(2\theta) &= 2\tfrac{2\xi(1-\xi^2)}{(1+\xi^2)^2},\\
\sin(\theta-\alpha) &= \tfrac{2\xi\cos\alpha - (1-\xi^2)\sin\alpha}{1+\xi^2}.
\end{align*}
Substituting,
\[\begin{split}
0 &= \tfrac{r_2}{r_1^2}(2\xi\cos\alpha - (1-\xi^2)\sin\alpha)(1+\xi^2) - 4\xi(1-\xi^2)\\
&= \tfrac{r_2}{r_1^2}(2\xi\cos\alpha+2\xi^3\cos\alpha - \sin\alpha + \xi^4\sin\alpha)\\
&\quad - 4\xi(1-\xi^2)\\
&= (\tfrac{r_2}{r_1^2}\sin\alpha)\xi^4 + (2\tfrac{r_2}{r_1^2}\cos\alpha+4)\xi^3\\
&\quad + (2\tfrac{r_2}{r_1^2}\cos\alpha-4)\xi - \tfrac{r_2}{r_1^2}\sin\alpha
.\end{split}
\]
This quartic equation 
can be solved analytically; 
the real root 
that corresponds 
to $\theta$ 
with the minimum $\fpfun{\ubar(\theta)}$ 
is used to generate 
$u_\pm = (c_0e^{\imath\theta} - \sbar)\phase{s}$, 
which is valid 
as long as $|u_\pm|^2 > y$.
Also, one must consider 
$\theta = \pm\pi$, 
which correspond to $\xi = \pm\infty$, 
in case either extreme point
minimizes $\fpfun{\ubar(\theta)}$.

\section*{Acknowledgments}

The authors would like to acknowledge 
Yoav Shechtman 
for insights relating to phase retrieval 
and coherent diffraction imaging, 
and for sharing image data,
and James Fienup 
for general discussions on phase retrieval. 

\ifCLASSOPTIONcaptionsoff
  \newpage
\fi

\IEEEtriggeratref{40}


\bibliographystyle{IEEEtran}
\bibliography{main}
%
%








\end{document}